# Nonreciprocal light transmission in parity-time-symmetric whispering-gallery microcavities

Bo Peng[1†], Şahin Kaya Özdemir[1†*], Fuchuan Lei[1,2], Faraz Monifi[1], Mariagiovanna Gianfreda[3,4], Gui Lu Long[2], Shanhui Fan[5], Franco Nori[6], Carl M. Bender[3], Lan Yang[1*]

[1]Department of Electrical and Systems Engineering, Washington University, St. Louis, MO 63130, USA

[2] State Key Laboratory of Low-dimensional Quantum Physics and Department of Physics, Tsinghua University, Beijing 100084, P.R. China

[3]Department of Physics, Washington University, St. Louis, MO 63130, USA

[4]Dipartimento di Matematica e Fisica Ennio De Giorgi, Universita del Salento and I.N.F.N. Sezione di Lecce, Via Arnesano, I-73100 Lecce, Italy.

[5]Department of Electrical Engineering, Stanford University, Stanford, California 94305, USA

[6]CEMS, RIKEN, Saitama 351-0198, Japan

*Correspondence to: yang@ese.wustl.edu, ozdemir@ese.wustl.edu

†These authors contributed equally to this work.

**Optical systems combining balanced loss and gain profiles provide a unique platform to implement classical analogues of quantum systems described by non-Hermitian parity-time- (PT-) symmetric Hamiltonians and to originate new synthetic materials with novel properties. To date, experimental works on PT-symmetric optical systems have been limited to waveguides in which resonances do not play a role. Here we report the first demonstration of PT-symmetry breaking in optical resonator systems by using two directly coupled on-chip optical whispering-gallery-mode (WGM) microtoroid silica resonators. Gain in one of the resonators is provided by optically pumping Erbium ($Er^{3+}$) ions embedded in the silica matrix; the other resonator exhibits passive loss. The coupling strength between the resonators is adjusted by using nanopositioning stages to tune their distance. We have observed reciprocal behavior of the PT-symmetric system in the linear regime, as well as a transition to nonreciprocity in the PT symmetry-breaking phase transition due to the significant enhancement of nonlinearity in the broken-symmetry phase. Our results represent a significant advance towards a new generation of synthetic optical systems enabling on-chip manipulation and control of light propagation.**

**Introduction.** Parity-time- (PT-) symmetric quantum Hamiltonian systems have attracted increasing attention during the past decade following the work of Bender and Boettcher *(1)*, who showed that the eigenvalues of non-Hermitian Hamiltonians $\hat{H}^\dagger \neq \hat{H}$ can still be entirely real if they respect PT-symmetry, $PT\hat{H} = \hat{H}PT$. It is now understood that one can interpret PT-symmetric systems as nonisolated physical systems having balanced absorption (loss) and amplification (gain). A remarkable feature of such systems is that they exhibit a phase transition *(1-3)* (*spontaneous PT-symmetry breaking*) if the parameter that controls the degree of non-Hermiticity exceeds a certain critical value. Beyond this critical threshold the spectrum is no longer real, eigenvalues become complex even though $PT\hat{H} = \hat{H}PT$ is still satisfied.

PT symmetry has been studied both experimentally [4-12] and theoretically [13-23] in a variety of physical systems, with experiments performed on electronic circuits [4], nuclear magnetic resonance [5], optics [6-8], metamaterials [9], microwave cavities [10], mechanical oscillators [11], and superconductors [12]. Optical systems have emerged as the most productive and versatile platform to study the fundamentals of PT symmetry and to test and explore PT-symmetric applications such as unidirectional invisibility and loss-induced transparency.

Here we report an experimental demonstration of a PT-symmetric system of two coupled whispering-gallery-mode (WGM) resonators. Our experiments differ from previous optics experiments in two significant ways. First, resonances play a dominant role in the dynamics and the evolution of our PT-symmetric system. Previous PT-symmetric optics experiments [6,7] used coupled waveguides in which the resonances play no role. Second, unlike previous experiments in which the coupling strength was fixed, our implementation enables us to control both the coupling strength and the amplification-to-absorption ratio, making it possible to probe phase transitions in the wider phase space of the two tunable parameters.

We have demonstrated reciprocal transport in the linear regime, in both the PT-symmetric unbroken and broken phases, and a significant enhancement of nonlinearity in the broken-symmetry regime, leading to a strong nonreciprocal transport in the nonlinear regime with very low power threshold, thereby providing a direct experimental clarification regarding the important issue of nonreciprocity in PT-symmetric systems. The realization of PT-symmetric systems with coexisting coherent perfect absorption (CPA) and lasing [21-23] require PT-symmetric optical resonators, which we are demonstrating for the first time here. Extending PT-symmetric concepts to optical resonators will open a fruitful research direction in which CPA-lasers and chaotic ray dynamics can be explored.

**<u>Whispering-gallery microcavities and experimental setup.</u>** In a WGM resonator light is confined by total internal reflection and circulates around the curved inner boundary of the resonator. WGM modes exhibit an evanescent tail that helps to couple light in and out of the resonator, and to probe the changes in or near the resonator for ultra-high performance sensing [24,25]. Moreover, the evanescent tail makes it possible to couple directly two or more WGM resonators. The performance of a WGM resonator is determined by its quality factor $Q$, which represents the total loss (material, radiation, scattering and coupling losses) experienced by the light in the mode.

Our system is composed of two directly coupled WGM microtoroidal resonators, each coupled to a different fiber-taper coupler (Fig. 1A-C). The first microtoroid ($\mu R_1$) is an active resonator made from $Er^{3+}$-doped silica formed using sol-gel synthesis [26-28] and the second microtoroid ($\mu R_2$) is a passive (no-gain-medium) resonator made from silica without dopants [26]. To have a controllable direct coupling between them, the microtoroids were fabricated at the edges of two separate chips. The chips are placed on nanopositioning systems to control precisely the distance and hence the coupling between the microtoroids. Optical gain in $\mu R_1$ is provided by optically pumping the $Er^{3+}$ ions, which emit in the 1550 nm band, with a pump light at 1460 nm band. The $Q$-factors of $\mu R_1$ and $\mu R_2$ in the 1550 nm wavelength band are $3.3 \times 10^6$ and $3 \times 10^7$, respectively, and $\mu R_1$ has a $Q$-factor of $2.4 \times 10^6$ in the spectral band of the pump (Fig. 1D).

As the pump laser power is increased, the gain starts to compensate the losses of $\mu R_1$ in the 1550nm band. This is reflected in the decreasing linewidth of the resonance line (Fig. 1E). When a weak probe light is coupled to $\mu R_1$ together with the pump light, we observe a strong resonance peak, indicating that the weak probe signal has been amplified by the gain provided by $Er^{3+}$ ions (Fig. 1F). The resonance wavelength of $\mu R_2$ is thermally tuned through the thermo-optic effect of silica. By controlling the detuning in resonant wavelengths between $\mu R_1$ and $\mu R_2$ we can mediate their coupling in the 1550 nm band. In addition, the coupling between the resonators is modified by changing the distance between them. There is no coupling between the resonators in the 1460 nm band; thus, the pump light exists only in $\mu R_1$. In all previous realizations of PT-symmetric photonic systems, the coupling strength was kept constant, while the gain-to-loss ratio was varied. The ability to control the amplification and absorption ratio and the coupling strength makes our platform highly versatile for investigating the novel physics of PT-symmetry.

**Parity-time symmetry breaking in WGM microcavities.** We conducted two sets of experiments using the apparatus described in Fig. 1. The first set of experiments determined the broken and unbroken PT phases as a function of the coupling strength. We studied the system using only the first waveguide (WG1) with ports 1 and 2 by moving the second waveguide (WG2) with ports 3 and 4 so far from the resonators that there was no coupling between WG2 and the resonators. The pump and the weak probe lasers were input at port 1 and the output transmission spectra were monitored at port 2 in the 1550 nm band. Without the pump, the coupled-resonator system acted as a passive photonic molecule characterized by two supermodes whose spectral distance increases with increasing coupling strength (that is, decreasing distance between µR$_1$ and µR$_2$) (Fig. 2A) [29,30]. This system became PT-symmetric when µR$_1$ was optically pumped to provide gain and µR$_2$ had a balanced loss. At fixed gain-loss ratio (i.e., pump power), we monitored the output port as a function of the coupling strength κ, observing that there was a threshold coupling strength $\kappa_{PT}$ at which the PT-symmetry phase transition occurred (Fig. 2 A&B). For $\kappa/\kappa_{PT} < 1$, the system is in the broken-symmetry phase, as seen in both the zero mode-splitting (Fig. 2A) and the nonzero linewidth difference (Fig. 2B). This indicates that the real parts of the eigenfrequencies have coalesced and that their imaginary parts are different. As $\kappa/\kappa_{PT}$ approaches 1 from below, the linewidth difference decreases and frequencies bifurcate (mode-splitting).

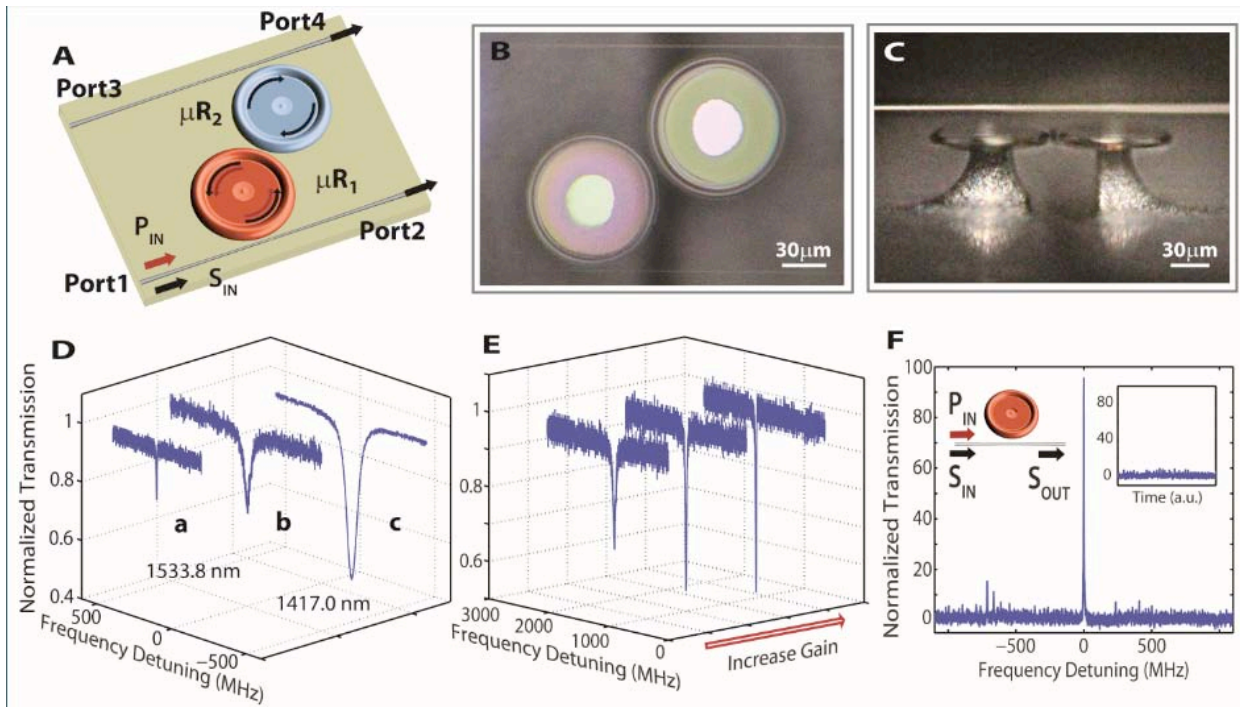

*Fig.1. Concept of the PT-symmetric WGM microcavities.* (A) The system consists of two directly coupled WGM resonators, and two fiber-taper waveguides. $\mu R_1$: active $Er^{3+}$-doped silica microtoroid. $\mu R_2$: passive silica microtoroid. $P_{IN}$: pump laser in 1460 nm band to excite $Er^{3+}$ ions which provide gain in the 1550 nm band. $S_{IN}$: probe light (signal) in the 1550 nm band. (B&C) Images of the top and side views of the coupled resonators. (D) Transmission spectra showing the resonance line of $\mu R_2$ at 1533.8 nm (a) and resonance lines of $\mu R_1$ at 1533.8 nm (b) and at 1417.0 nm (c). (E) Gain provided by $Er^{3+}$-ions in $\mu R_1$ leads to narrower and deeper resonance lines as the pump power (gain) is increased, implying an increasing Q-factor. (F) Weak probe light is amplified when it is coupled to $\mu R_1$ together with the pump light. Inset shows that without the weak $S_{IN}$ there is no resonance enhancement.

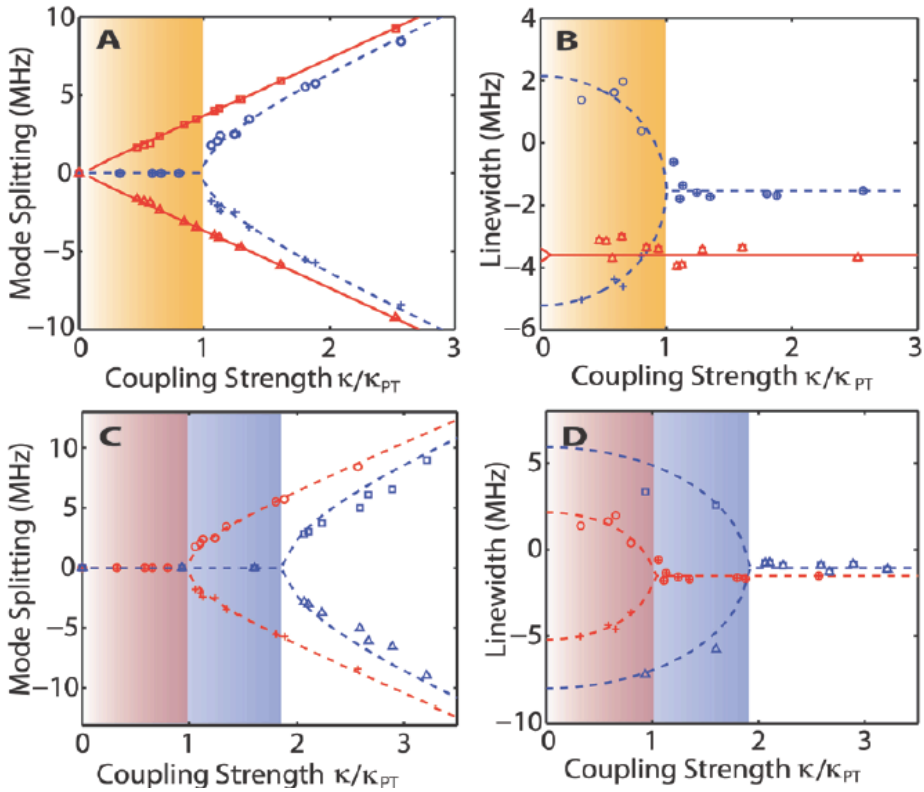

*Fig.2. Experimentally observed mode-splitting and linewidth-difference of the supermodes in the coupled-microresonator system as a function of coupling strength κ. Mode-splitting (A&C) and linewidth (B&D) variation correspond to the difference between the real parts and changes in the imaginary parts of the eigenfrequencies, respectively. (A&B) Comparison when resonators are passive (no gain in µR$_1$: red square and triangular marks) with Q factors $2.9\times10^7$ and $3.0\times10^7$, respectively, for µR$_1$ and µR$_2$, and when one resonator is active and the other is passive (with gain in µR$_1$: blue circular and cross marks). (C&D) Effect of the initial Q-factor (loss) of µR$_2$ on the eigenfrequencies. Two resonance modes with Q-factors $2.0\times10^7$ (blue) and $3.0\times10^7$ (red) are chosen for µR$_2$. Shaded regions correspond to the broken-PT-symmetric region when gain and loss are balanced.*

Next, we chose two different WGMs with $Q$-factors $2.0\times10^7$ and $3.0\times10^7$ in $\mu R_2$ and adjusted the pump power so that loss-gain ratio was nearly balanced. We observed that the transition from the broken to unbroken phase occurs at different coupling strengths for modes with different $Q$, that is, different initial loss (Fig. 2C&D). The lower the $Q$-factor, the higher the $\kappa_{PT}$ for a PT phase transition. Typical transmission spectra in broken- and unbroken-PT-symmetric phases are given in Fig. 3.

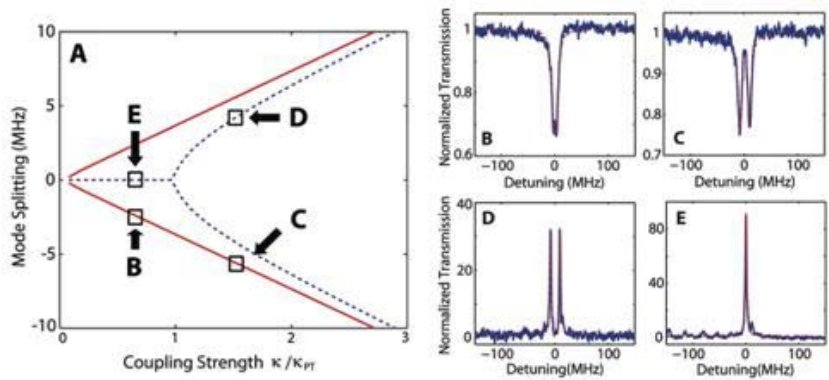

*Fig.3. Experimentally obtained transmission spectra in broken- and unbroken-PT-symmetric regions of Fig. 2A. (A) Mode-splitting determined by the difference between the real parts of the eigenfrequencies of the supermodes of the coupled WGM-resonators system as a function of the coupling strength. The transmission spectra at different coupling condition in broken- and unbroken-symmetry regions are given in panels (B)-(E). The red straight lines depict the case with two passive resonators. The dotted blue lines depict the case in which one resonator is active and the other is passive with balanced loss and gain. (B) Both resonators are passive (no gain) and the coupling is small. Splitting is barely seen. (C) Both resonators are passive and the coupling is strong. Mode-splitting is clearly seen. (D) Coupled passive and active resonators in the unbroken-symmetry region. Split resonance peaks are clearly seen. (E) Coupled passive and active resonators in the broken-symmetry region.*

The PT phase transition can be understood intuitively as follows. If the coupling between the resonators is weak, the energy in the active resonator cannot flow fast enough into the passive resonator to compensate for the absorption. Thus, the system cannot be in equilibrium and the eigenfrequencies are complex, implying exponential growth or decay. However, if the coupling strength exceeds a critical value, then the system can attain equilibrium because the energy in the active resonator can flow rapidly enough into the passive one to compensate the dissipation.

In our experiments the frequency bifurcation (splitting) is not in orthogonal directions (Fig. 2) as would be expected for ideal systems with exactly balanced gain and loss. Instead, the bifurcation is smooth and the degree of smoothness (how much the system deviates from the exactly balanced case) depends on the pump power. In order to understand the origin of this behavior, we have revisited the equations of motion for coupled oscillators, which show that, for unbalanced gain and loss, the eigenfrequencies are never exactly real. Instead, there is a region of κ where the difference in imaginary parts (linewidth difference) is large but the difference in real parts (mode splitting) is small (but nonzero). There is a second region where the linewidth difference is small but nonzero and the mode splitting is large. In practical implementations it is impossible to equalize the loss and gain exactly, so the mathematical prediction of a smooth bifurcation is physically realistic and is consistent with our experimental observations (Fig. 2).

**Nonreciprocal light transmission: Optical diode action in PT-symmetric WGM microcavities.** As predicted theoretically and demonstrated experimentally, PT-symmetric systems exhibit distinct behaviors including unidirectional or asymmetric transmission and invisibility, and enhanced or reduced reflections in the broken-PT-symmetric regime, where propagation is not invariant under the exchange of loss and gain. In the second set of experiments we demonstrate for the first time nonreciprocal behavior in PT-symmetric system in the optical frequency range, which allows light to pass in only one direction. Strikingly, this strong nonreciprocal light transmission is associated with nonlinearity enhancement due to strong field localization induced by PT-symmetry-breaking. In the linear regime, light transmission is still reciprocal. It's worth noting here that asymmetric behavior of PT symmetry was previously shown in electronic circuits at a frequency of 30 kHz [4]; however, a signal in kHz differs from the optical frequency in our work by almost ten orders of magnitude (kHz electronic signal versus THz optical signal), and the microscopic physics of experiments in electronic and optic domains are completely different.

It is known theoretically that a linear static dielectric system, even in the presence of gain and loss, cannot have a nonreciprocal response [17,18,31-34]. On the other hand, a system with nonlinearity can exhibit very strong nonreciprocity. We tested this in our PT-symmetric system shown in Fig. 1A where the transmission from input port 1 (4) to the output port 4 (1) is defined as the forward $T_{1\to4}$ (backward $T_{4\to1}$). We first monitored the output spectra at port 1 as the power of the input probe at port 4 was varied when the system was in the broken- or unbroken-symmetry phases. A clear nonlinear response was observed in the symmetry-broken phase, in contrast to the linear response in the unbroken phase (Fig. 4A). At low power levels, where the input-output relation was linear, the system was reciprocal in both the broken- and unbroken-symmetry regions (Fig. 4B&C). Thus we have provided a direct experimental clarification of the issue of reciprocity in PT symmetric systems; that is, PT symmetry or PT-symmetry breaking alone is not sufficient for nonreciprocal light transmission.

As we increased the input power, the system remained in the linear regime for the unbroken-symmetry phase, whereas the input-output relation becomes nonlinear in the broken phase (Fig. 4A). These results indicate nonlinearity enhancement (i.e., the threshold for nonlinearity is lower) in the broken-symmetry phase, due to the stronger field localization into the resonator with gain [17], as compared to the unbroken-symmetry phase.

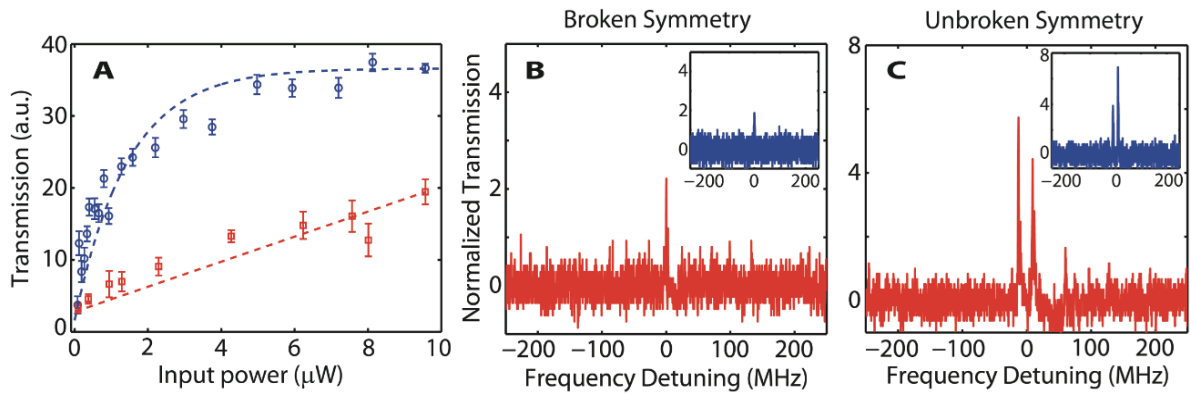

*Fig.4. Input-output relation in PT-symmetric WGM resonators and reciprocity in the linear regime. (A) Input-output relation is linear in the unbroken-symmetry region (red square symbols) and nonlinear in the broken-symmetry region (blue circle symbols). Transmission spectra were obtained at port 1 when the input was at port 4. (B&C) Transmission spectra in the linear regime (low input power levels) show reciprocal light transmission at forward (blue inset) and backward (red spectra) directions in both the broken- (B) and unbroken- (C) symmetry regions.*

Due to the stronger nonlinearity in the broken-symmetry case, the PT-symmetry transition is associated with a transition from reciprocal to nonreciprocal behavior. When the pump at port 1 is OFF ($\mu R_1$ and $\mu R_2$ are passive) and a weak probe light is input at ports 1 or 4, we observe resonance peaks in the forward or backward transmissions (Fig. 5A(a) & 5B(a)) with no resolvable mode splitting. When the pump is set ON and the gain and loss are balanced as much as possible so as to operate the system in the unbroken-PT-symmetric region, transmission spectra showing amplified signals with clearly resolved split peaks are observed at the outputs in this strong-coupling region (Fig. 5A(b) & 5B(b)). However, when the coupling strength is decreased so that the system transits into the broken-symmetry region, forward transmission reduces to zero $T_{1\to 4} \sim 0$ (Fig. 5A(c)) but the backward transmission remains high (Fig. 5B(c)). The transmission spectra show a single resonance peak, as expected from the theory. Thus, in the broken-symmetry region, the input at port 4 is transmitted to port 1 at resonance; however the input at port 1 cannot be transmitted to port 4, in stark contrast with what is observed for the unbroken-symmetry region. This indicates nonreciprocal light transport between ports 1&4.

Unlike previous experiments demonstrating nonreciprocal transport in non-PT microresonators [34,35] and asymmetric behavior in PT-electronics [4], here we observe a complete absence of resonance peak in the forward transmission. Also, the transmitted signal here is not from spontaneous emission of the gain medium. Without the weak injected signal at the input port 4, the output at port 1 is at the noise level, and no resonance peak is observed. Only with the weak signal present, resonance enhancement and thus the peak is observed (Fig. 5B(b)&(c) inset).

The nonreciprocity is observed regardless of whether the pump is input at port 1 or 2. Note that the pump cannot be at port 3 or 4 because the absence of coupling between $\mu R_1$ and $\mu R_2$ at the pump band prevents the optical pumping of $Er^{3+}$ ions in $\mu R_1$. We have observed similar nonreciprocity between ports 2&3. These results imply that PT-symmetric WGM microcavities in the broken-symmetry region can give rise strong nonreciprocal effects due to nonlinearity

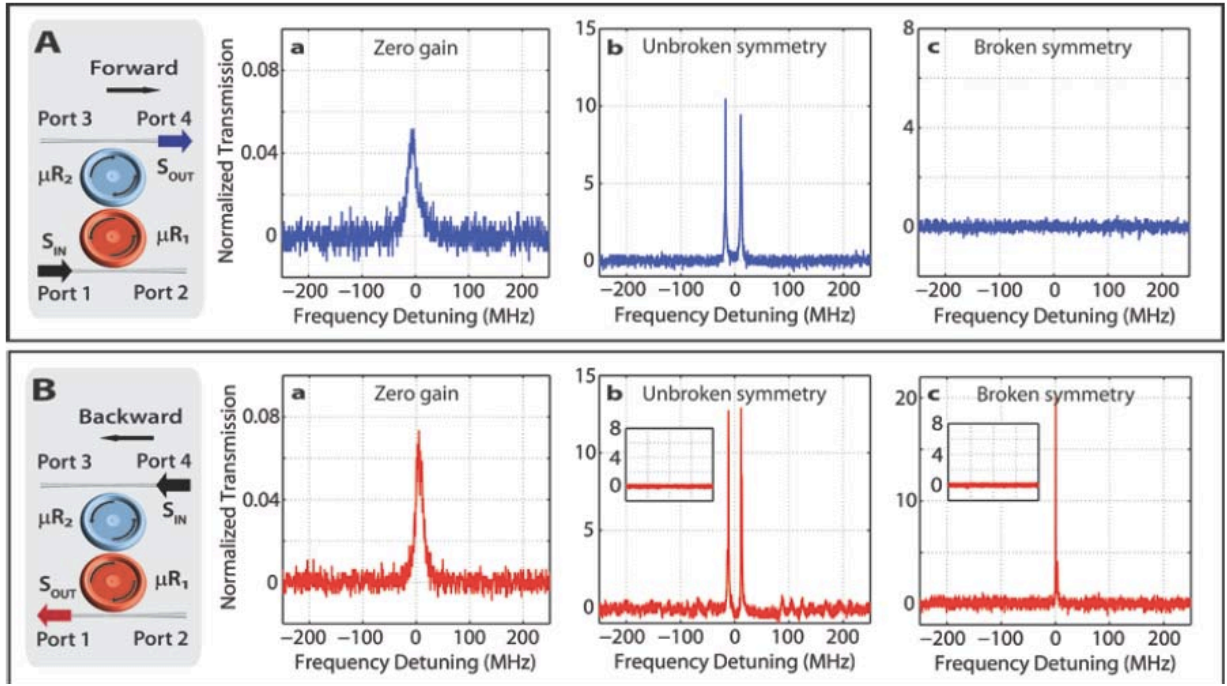

*Fig.5. Experimentally observed unidirectional transmission for PT-symmetric WGM microresonators in the nonlinear regime.* When both resonators are passive (no gain), the transmission is bi-directional (reciprocal), and light is transmitted in both forward (*A(a)*) and backward directions (*B(a)*). In the unbroken-symmetry region, where the coupling exceeds the critical value and gain and loss are balanced, the transmission is still bi-directional (*A(b)* & *B(b)*). Mode splitting due to coupling is now resolved because gain compensates loss leading to narrower linewidths. In the broken-symmetry region (*A(c)* & *B(c)*), transmission becomes unidirectional (nonreciprocal). Input in the backward direction reaches the output (*B(c)*), but input in the forward direction does not (*A(c)*). This resembles the action of a diode and implies that an all-optical on-chip diode with PT-symmetric WGM microcavities operates in the broken-symmetry region. Inset in (*B(b)*) and (*B(c)*) shows the signal at port 1 when there is no input signal at port 4.

**Conclusion and summary.** In summary, our experiments extend the concept of PT-symmetric optics and the PT-symmetric nonreciprocity (diode action) from centimeter- and meter-scale structures to the domain of on-chip micro-scale structures, and more importantly from waveguides to microresonators. This work significantly contributes to the field in many ways. First, it presents the first PT-symmetric optical microresonators with a clear demonstration of PT symmetry breaking. Second, it provides an experimental demonstration of the enhancement of nonlinearity (reduction of threshold for nonlinearity) in the broken-symmetry phase. Third, it presents an experimental verification that PT-symmetry alone is not sufficient to obtain nonreciprocal behavior; operation in the nonlinear regime is also necessary. A PT-symmetric system will always be reciprocal in the linear regime, regardless of whether PT-symmetry is broken or unbroken. Thus, this work provides a direct experimental proof of the issue of reciprocity in PT-symmetric systems. Fourth, it provides for the first time an experimental demonstration of nonreciprocal light transmission -- without magneto-optic effect -- in a PT-symmetric system operated in the nonlinear regime and broken-symmetry phase.

Although we used fiber-taper coupled microtoroids, our techniques can be extended to other WGM resonators with integrated coupling waveguides and to photonic crystal cavities. Similarly, gain could be provided by quantum dots or other rare-earth-ions and also through nonlinear processes, such as Raman or parametric gain. Like any nonreciprocal device utilizing resonant effects, our PT-symmetric all-optical diode is bandwidth-limited [34,35]. However, by thermally tuning the resonance wavelengths and by using active resonators doped with multiple rare-earth-ions, operation over large-wavelength bands should be possible. Coupled WGM microresonators provide a comprehensive framework for understanding resonance effects in PT-symmetric optical systems, whose optical properties are unconventional, thereby aiding the development of CPA lasers and on-chip synthetic structures to harness the flow of light.

**References and Notes:**
1. S. Boettcher, C. M. Bender, Real spectra in non-Hermitian Hamiltonians having PT symmetry. *Phys. Rev. Lett.* **80**, 5243 (1998).
2. C. M. Bender, Making sense of non-Hermitian Hamiltonians. *Rep. Prog. Phys.* **70**, 947 (2007).
3. A. Mostafazadeh, Pseudo-Hermiticity versus PT symmetry: the necessary condition for the reality of the spectrum of a non-Hermitian Hamiltonian. *J. Math. Phys.* **43**, 205 (2002).
4. N. Bender *et al.*, Observation of Asymmetric Transport in Structures with Active Nonlinearities. *Phys. Rev. Lett.* **110**, 234101 (2013).
5. C. Zheng, L. Hao, G. L. Long, Observation of fast evolution in parity-time-symmetric system. *Phil. Trans. R. Soc. A*. **371**, 20120053 (2013).
6. C. E. Rüter *et al.*, Observation of parity-time symmetry in optics. *Nature Phys.* **6**, 192 (2010).
7. A. Regensburger *et al.*, Parity-time synthetic photonic lattices. *Nature* **488**, 167 (2012).
8. L. Feng *et al.* Nonreciprocal light propagation in a silicon photonic circuit. *Science* **333**, 729 (2011).
9. L. Feng *et al.*, Experimental demonstration of a unidirectional reflectionless parity-time metamaterial at optical frequencies. *Nature Mat.* **12**, 108 (2012).

10. S. Bittner *et al.*, PT symmetry and spontaneous symmetry breaking in a microwave billiard. *Phys. Rev. Lett.* **108**, 024101 (2012).

11. C. M. Bender, B. K. Berntson, D. Parker, E. Samuel, Observation of PT phase transition in a simple mechanical system. *Am. J. of Phys.* **81**, 173 (2013).

12. N. M. Chtchelkatchev, A. A. Golubov, T. I. Baturina, V. M. Vinokur, Stimulation of the fluctuation superconductivity by PT symmetry. *Phys. Rev. Lett.* **109**, 150405 (2012).

13. C. Hang, G. Huang, V. V. Konotop, PT symmetry with a system of three-level atoms. *Phys. Rev. Lett.* **110**, 083604 (2013).

14. E.-M. Graefe, Stationary states of a PT symmetric two-mode Bose-Einstein condensate. *J. Phys. A: Math. Theor.* **45** 444015 (2012).

15. G. S. Agarwal, K. Qu, Spontaneous generation of photons in transmission of quantum fields in PT-symmetric optical systems. *Phys. Rev. A* **85**, 031802 (2012).

16. X Zhu, L. Feng, P. Zhang, X. Yin, and X. Zhang, One-way invisible cloak using parity-time symmetric transformation optics. *Opt. Lett.*, **38**, 2821 (2013).

17. A. A. Sukhorukov, Z. Xu, and Y.. S. Kivshar, Nonlinear suppression of time reversals in PT-symmetric optical couplers. *Phys. Rev. A* **82**, 043818 (2010).

18. H. Ramezani, T. Kottos, R. El-Ganainy, D. N. Christodoulides. Unidirectional nonlinear PT-symmetric optical structures. *Phys. Rev. A*. **82**, 043803 (2010).

19. H. Benisty *et al.*, Implementation of PT symmetric devices using plasmonics: principle and applications. *Opt. Exp.* **19**, 18004 (2011).

20. N. Lazarides, G. P. Tsironis, Gain-driven discrete breathers in PT-symmetric nonlinear metamaterials. *Phys. Rev. Lett.* **110**, 053901 (2013).

21. S. Longhi, PT-symmetric laser absorber. *Phys. Rev. A* **82**, 031801(R) (2010).

22. Y. D. Chong, L. Ge, A. D. Stone, PT-symmetry breaking and laser-absorber modes in optical scattering systems. *Phys. Rev. Lett.* **106**, 093902 (2011).

23. M. Liertzer *et al.*, Pump-induced exceptional points in lasers. *Phys. Rev. Lett.* **108**, 173901 (2012).

24. L. He, S. K. Ozdemir, J. Zhu, W. Kim, L. Yang, Detecting single viruses and nanoparticles using whispering gallery microlasers. *Nature Nanotech.* **6**, 428 (2011).

25. J. Zhu *et al.*, Single nanoparticle detection and sizing by mode-splitting in an ultra-high-*Q* microtoroid resonator. *Nature Photon.* **4**, 46 (2010).

26. D. K. Armani, T. J. Kippenberg, S. M. Spillane, K. J. Vahala, Ultra-high-*Q* toroid microcavity on a chip. *Nature* **421**, 925 (2003).

27. L. Yang, T. Carmon, B. Min, S. M. Spillane, K. J. Vahala, Erbium-doped and Raman microlasers on a silicon chip fabricated by the sol-gel process. *App. Phy. Lett*., **86**, 091114 (2005).

28. L. He, S. K. Ozdemir, L. Yang, Whispering gallery microcavity lasers. *Laser &Photon. Rev.* **7**, 60 (2013).

29. B. Peng, S. K. Ozdemir, J. Zhu, L. Yang, Photonic molecules formed by coupled hybrid resonators. *Opt. Lett.* **37**, 3435 (2012).

30. S. K. Ozdemir, J. Zhu, L. He, L. Yang, Estimation of Purcell factor from mode-splitting spectra in an optical microcavity. *Phys. Rev. A.* **83**, 033817 (2011).

31. S. Fan *et al.*, Comment on "Nonreciprocal Light Propagation in a Silicon Photonic Circuit" *Science* **335**, 38 (2012).

32. X. Yin and X. Zhang, Unidirectional light propagation at exceptional points, *Nature Mat.* **12**, 175 (2013).

33. Z. Yu, S. Fan, Complete optical isolation created by indirect interband photonic transitions. *Nature Photon.* **3**, 91 (2009).

34. L. Fan *et al.* An all-silicon passive optical diode. *Science* **335,** 447 (2012).

35. L. Bi *et al.* On-Chip Optical Isolation in Monolithically Integrated Nonreciprocal Optical Resonators. *Nature Photon.* **5,** 12 (2011).

**Acknowledgments:** This work is partially supported by ARO grant No. W911NF-12-1-0026. CMB thanks US DoE. SKO and LY conceived the idea and designed the experiments; BP performed the experiments with help from FL, FM and SKO. Theoretical background and simulations were provided by FL, FM, MG, CMB, SF and FN. All authors discussed the results, and SKO and LY wrote the manuscript with inputs from all authors. LY supervised the project. The authors thank Xu Yang for her help.